\documentstyle[epsfig,preprint,aps]{revtex}
\begin{document}
\tighten

\def\bea{\begin{eqnarray}}
\def\eea{\end{eqnarray}}
\def\beas{\begin{eqnarray*}}
\def\eeas{\end{eqnarray*}}
\def\nn{\nonumber}
\def\ni{\noindent}
\def\G{\Gamma}
\def\F{\Phi}
\def\P{\Pi}
\def\D{\Delta}
\def\d{\delta}
\def\l{\lambda}
\def\L{\Lambda}
\def\g{\gamma}
\def\m{\mu}
\def\n{\nu}
\def\s{\sigma}
\def\tt{\theta}
\def\b{\beta}
\def\a{\alpha}
\def\f{\phi}
\def\fh{\phi}
\def\y{\psi}
\def\z{\zeta}
\def\p{\pi}
\def\r{\rho}
\def\e{\epsilon}
\def\ve{\varepsilon}
\def\ch{{\cal H}}
\def\cc{{\cal C}}
\def\cd{{\cal D}}
\def\cf{{\cal F}}
\def\cg{{\cal G}}
\def\cl{{\cal L}}
\def\cv{{\cal V}}
\def\cz{{\cal Z}}
\def\co{{\cal O}}
\def\pl{\partial}
\def\ov{\over}
\def\~{\tilde}
\def\rar{\rightarrow}
\def\lar{\leftarrow}
\def\lrar{\leftrightarrow}
\def\rra{\longrightarrow}
\def\lla{\longleftarrow}
\def\8{\infty}
\def\h{\hbar}
\def\yb{\bar{\psi}}
\def\lg{\langle}
\def\rg{\rangle}
\def\pls{\partial\!\!\!/}
\def\bs{b\!\!\!/}
\def\ps{p\!\!\!/}
\def\as{A\!\!\!/}
\def\Jy{\sinh(eFs)}
\def\V{(F\s b-\s Fb)}
\def\Vf{(F\s b-\s Fb)\g_5}
\def\U{(b\s F-bF \s)}
\def\edt{\end{document}}


\title{Induced Lorentz- and CPT-violating Chern-Simons term
in QED: Fock-Schwinger proper time method}

\author{J.~-M. Chung$^a$\footnote
  {Electronic address: jmchung@apctp.org}
and B.~K. Chung$^{a,b}$\footnote
  {Electronic address: bkchung@khu.ac.kr}}
\address{$^a$Asia Pacific Center for Theoretical Physics, 
Seoul 130-012, Korea\\
$^b$Research Institute for Basic Sciences
and Department of Physics\\ Kyung Hee University, Seoul 130-701, Korea\\
{~}}
                                  
\maketitle \draft                                     
\begin{tighten}
\begin{abstract}
\indent
Using the Fock-Schwinger proper time method, 
we calculate the induced Chern-Simons term arising from 
the Lorentz- and CPT-violating sector of quantum electrodynamics
with a $b_\m \yb \g^\m \g_5 \y$ term. Our result to all orders
in $b$ coincides with a recent linear-in-$b$ calculation by
Chaichian et al. [hep-th/0010129 v2].
The coincidence was pointed out by Chung 
[Phys. Lett. {\bf B461} (1999) 138] and P\'{e}rez-Victoria 
[Phys. Rev. Lett. {\bf 83} (1999) 2518] in the
standard Feynman diagram calculation
with the nonperturbative-in-$b$ propagator.
\end{abstract}
\end{tighten}

The Lorentz- and CPT-violating Chern-Simons modification to the Maxwell 
theory was first proposed a decade ago \cite{cfj}. An important feature of 
the Chern-Simons term \cite{djt} is that Lagrange density is not gauge 
invariant,
but the action and equations of motion are gauge invariant. In their 
Lorentz violating extension of the standard model, Colladay and  
Kosteleck\'y posed the question whether such a term is induced 
when the Lorentz- and 
CPT-violating term $\yb \bs\g_5 \y$ ($b_\m$ is a constant four vector)
is added to the conventional Lagrangian of QED \cite{ck2}.

Recently, several calculations have been carried out 
to determine the radiatively induced
Chern-Simons term from the Lorentz- and CPT-violating 
fermion sector
\cite{ck2,jm1,jm2,jm3,pv,jk,ch,chan,ht,ccg,nrs,bsn,bn}:
 \bea
 \cl=\yb(i\pls-e\as-\g_5\bs-m)\y\;.         \label{fms}
 \eea
Jackiw and Kosteleck\'y clarified that the induced Chern-Simons term 
is not uniquely determined \cite{jk}. It depends on whether one uses 
a nonperturbative definition or a perturbative definition of 
the theory defined by Eq.~(\ref{fms});\footnote{In the  
nonperturbative definition, we use the $b$-exact propagator
$S(p)={i\ov \ps-m-\g_5\bs}$ for the standard perturbation calculation.
(See the work of Colladay and Kosteleck\'y \cite{ck1} for detailed analysis 
of the extended Dirac theory with this propagator.)
In the perturbative definition, the propagator has the  $b$-independent
form $S(p)={i\ov \ps-m}$ and $-\g_5\bs$  is considered as 
an interaction vertex.}
In a nonperturbative formalism, radiative 
corrections induce a definite nonzero Chern-Simons term, while when 
a perturbative formalism is used, radiative corrections are finite but
undetermined. The regularization scheme one chooses to adopt can generate
further ambiguity in both nonperturbative and perturbative formalisms.

In the standard Feynman 
diagram calculation with the nonperturbative-in-$b$ propagator,
Chung \cite{jm3} and P\'{e}rez-Victoria \cite{pv} 
demonstrated that the result to all orders in $b$ coincides
with the previous linear-in-$b$ calculation by Chung and Oh \cite{jm1}
as well as Jackiw and Kosteleck\'y \cite{jk}. 
More recently, nonstandard approaches were employed also in the calculation 
of the induced Chern-Simons term. Chan \cite{chan} and 
Chaichian et al.\cite{ccg},
used the covariant derivative expansion \cite{cgd} 
and the Fock-Schwinger proper time method \cite{fk,sc}, 
respectively, to obtain the induced Chern-Simons term. 
The common feature in these two nonstandard 
approaches is to develop a series of local effective Lagrangian in powers 
of $\P_\m= i\pl_\m-eA_\m$, rather than in powers of $i\pl_\m$  
and $A_\m$ separately.

The purpose of this work is to calculate, using the Fock-Schwinger proper 
time method, the induced Chern-Simons term 
arising from the Lorentz- and CPT-violating sector of QED
with a $\yb \bs\g_5 \y$ term keeping the full $b$ dependence
in order to see whether the coincidence of all-order-in-$b$ result with 
linear-in-$b$ calculation takes place in this nonstandard approach.

The effective action, $\G_{\rm eff}$, of the theory 
defined by Eq.~(\ref{fms}) is given by
 \bea
 \G_{\rm eff}=-i\ln
 \biggl(\int \cd \yb \cd \y \,{\rm e}^{i\int d^4x \cl}\biggr)
 =-i{\rm Tr}\ln(i\pls-e\as-\g_5\bs-m)\;, \label{ea}
 \eea
where the trace ``Tr'' is taken over both spinor indices 
(of any combinations of $4 \times 4$ Dirac matrices) and 
space-time coordinates.

Let us decompose the trace in 
Eq.~(\ref{ea}) in the following manner, as was done in Eq.~(3) of 
Ref.~\cite{ccg}:
 \bea
 {\rm Tr}\ln(i\pls-e\as-\g_5\bs-m)&=&{\rm Tr}\ln(i\pls-e\as-m)\nn\\
 &&+\int_0^1 dz\, 
 {\rm Tr}\biggl({1\ov -i\pls+e\as+z\g_5\bs+m}\g_5\bs\biggr)\;. 
 \label{decomp}
 \eea
Then, the effective action can be written down as follows: 
 \beas
 \G_{\rm eff}=\G_{\rm eff}^{(0)}+\G_{\rm eff}^{(1)}\;,
 \eeas
where
 \bea
 \G_{\rm eff}^{(0)}&=&-i{\rm Tr}\ln(i\pls-e\as-m)\;,\nn\\
 \G_{\rm eff}^{(1)}&=&-i\int_0^1 dz\, 
 {\rm Tr}\biggl({1\ov -i\pls+e\as+z\g_5\bs+m}\g_5\bs\biggr)\;.
 \label{defg}
 \eea
The induced Chern-Simons term is contained in  $\G_{\rm eff}^{(1)}$.
By neglecting the dependence of $b$ in the denominator of the trace 
in  $\G_{\rm eff}^{(1)}$, the authors of Ref.~\cite{ccg} obtained 
the {\rm undetermined} induced Chern-Simons 
action\footnote{This undeterminicity arises from an intrinsic 
ambiguity in the limit $\lim_{x\rar 0}x_\m x_\n/x^2$. 
(See Eq.~(\ref{cgmn}) below.)  
This limit has directional dependence in a strict mathematical 
sense as emphasized in Ref.~\cite{ccg}.}
\beas
\G_{\rm CS}^{(1)}={ce^2\ov 4\p^2}\int d^4 x \,
\e^{\m\n\l\r}b_\m A_\n F_{\l\r}\;. 
~~~(\mbox{$c$ being an arbitrary contant})
\eeas
If one keeps the full dependence of $b$ in the calculation of
$\G_{\rm eff}^{(1)}$ for extracting the induced Chern-Simons term, 
one expects that the induced Chern-Simons action would take the following 
form:   
\bea
\G_{\rm CS}^{(1)}={ce^2\ov 4\p^2}
\biggl[1+f\biggl({b^2\ov m^2}\biggr)\biggr]
\int d^4 x \,\e^{\m\n\l\r}b_\m A_\n F_{\l\r}\;, \label{f}
\eea
where $f(b^2/m^2)$ is some function of a single argument 
$b^2/m^2$. In what follows, we are to show by explicit calculation 
that this function $f(b^2/m^2)$, indeed, vanishes. 

Now, let us introduce a (fermionic) Green's function $G(x,x')$ as the 
inverse of the operator $-i\pls+e\as(x)+z\g_5\bs+m$, i.e., 
let us assume 
that $G(x,x')$ satisfies the following inhomogeneous differential
equation
 \bea
 [-i\pls+e\as(x)+z\g_5\bs+m]G(x,x')=\d^4(x-x')\;.  \label{fp}
 \eea
Then, $\G_{\rm eff}^{(1)}$ of Eq.~(\ref{defg}) can be written as  
 \bea
 \G_{\rm eff}^{(1)} =\int d^4 x \cl_{\rm eff}^{(1)} 
 =-i\int d^4 x\int_0^1 dz\,
 {\rm tr}[ G(x,x')\g_5\bs]_{x'\rar x}\;.\label{g1}
 \eea
The trace ``tr'' in this equation is taken over the spinor indices 
only and the limit ${x'\rar x}$ is taken by averaging the forms obtained 
by letting $x'$ approach from the future and from the past \cite{sc}.
Further, if one introduces a (bosonic) Green's function 
$\D$ as follows:
 \bea
 G(x,x')=[i\pls-e\as(x)-z\g_5\bs+m]\D(x,x')\;, \label{gd1}
 \eea
or, equivalently,
 \bea
 G(x,x')=[-i\pl'_\m-eA_\m(x')]\D(x,x')\g^\m
 +\D(x,x')[-z\g_5\bs+m]\;, \label{gd2}
 \eea
then, one finds that Eq.~(\ref{fp}) becomes
 \bea
 \ch\D(x,x')=\d^4(x-x')\;,\label{h1}
 \eea
where  $\ch$ is defined as
 \bea
 \ch=-\P^2+m^2+z^2b^2+{e\ov 2}\s_{\m\n}F^{\m\n}
 +2iz\s^{\m\n}\P_\m b_\n\g_5\;, \label{h2}
 \eea
with  $\P_\m\equiv i\pl_\m-eA_\m$.
Borrowing the quantum mechanical 
matrix element notation, the  Green's function $\D(x',x'')$ can be expressed
as follows:
 \bea
 \D(x',x'')=\lg x'|\ch^{-1}|x''\rg &=& 
 i\int_0^\8 ds\,\lg x'|e^{-i\ch s}|x''\rg\;.\label{d1}
 \eea
In the $s$ integration, a convergence factor $-i\e$ ($\e>0$)
in $\ch$ is understood. 
The idea of the Fock-Schwinger proper time method  is to consider $\ch$
in Eqs.~(\ref{h1}) and (\ref{h2}) as a Hamiltonian that governs the
evolution in ``time'' $s$ 
of a hypothetical quantum mechanical system. Then the integrand in 
Eq.~(\ref{d1}) becomes a transition amplitude $\lg x'(s)|x''(0)\rg$.
Thus, the solution $\D(x',x'')$  can be written,  
in terms of this transition amplitude, as follows:
 \bea
 \D(x',x'')=i\int_0^\8 ds\,\lg x'(s)|x''(0)\rg\;.\label{d2}
 \eea

Now, let us evaluate the transition amplitude $\lg x'(s)|x''(0)\rg$.
It satisfies the evolution equation
 \bea
 i\pl_s\lg x'(s)|x''(0)\rg=\lg x'(s)|\ch\bbox( x(s),
 \P(s)\bbox)|x''(0)\rg\;,   \label{ev}
 \eea
with the boundary condition
 \bea
 \lg x'(s)|x''(0)\rg_{s\rar 0}=\d^4(x'-x'')\;. \label{bc}
 \eea
In order to obtain the Hamiltonian in the Heisenberg representation,
$\ch\bbox( x(s), \P(s)\bbox)$, one first has to know the 
evolutions of $x$ and $\P$. Using the commutation relations  
 \beas
 [x^\m,\P^\n]=-ig^{\m\n}\;,~~~
 [\P^\m,\P^\n]=-ieF^{\m\n}\;,
 \eeas
one can obtain the equations of motion for the operators $x(s)$ and
$\P(s)$:
 \bea
 \dot{x}_\m=i[\ch,x_\m]&=&2\P_\m-2iz\s_{\m\n}b^\n\g_5\;,\nn\\
 \dot{\P}_\m=i[\ch,\P_\m]&=&2eF_{\m\n}\P^\n+ie\pl^\n F_{\m\n}
 +{e\ov 2}\pl_\m F_{\r\n}\s^{\r\n}-2izeF_{\m\r}\s^{\r\n}b_\n\g_5\;.
 \label{xp}
 \eea
In the constant $F_{\m\n}$ approximation, these equations are solved,
in matrix notation, as follows:
 \bea
 \P(s)&=&{\rm e}^{2e F s}(\P(0)-iz\s b\g_5)+iz\s b\g_5\;,\nn\\
 x(s)&=&x(0)+(e F)^{-1}\Bigl({\rm e}^{2e F s}-1\Bigr)(\P(0)-iz\s b\g_5)\;.
 \eea
From these two relations, one can find  
 \bea
 \P(s)&=&{1\ov 2}eF{\rm e}^{e F s}[\sinh(e F s)]^{-1}[x(s)-x(0)]
 +iz\s b\g_5\;,\nn\\
 \P(0)&=&{1\ov 2}eF{\rm e}^{-e F s}[\sinh(e F s)]^{-1}[x(s)-x(0)]
 +iz\s b\g_5\;,\nn\\
 \P^2(s)&=&[x(s)-x(0)]K[x(s)-x(0)]\nn\\   
 &&+ize[x(s)-x(0)]{\rm e}^{-e F s}[\sinh(e F s)]^{-1}F\s b\g_5-3z^2b^2\;,
 \label{pppp}
 \eea
where
 \beas
 K={1\ov 4}e^2F^2[\sinh(e F s)]^{-2}\;.
 \eeas
Further, using the commutation relation
 \beas
 [x_\m(s),x_\n(0)]=
 i\Bigl[(eF)^{-1}({\rm e}^{2 e F s}-1)\Bigr]_{\m\n}\;,
 \eeas
one can find\footnote{The trace denoted by ``TR'' extends over
Lorentz indices only: ${\rm TR}(AB)=A_{\m\n}B^{\n\m}$. 
In particular, note that the
expression ${\rm TR}(\s F)=\s_{\m\n}F^{\n\m}$ is still a matrix
valued quantity because each $\s_{\m\n}$ is a $4\times 4$ matrix.}
 \bea
 x(0)K x(s)=x(s) K x(0)-{i\ov 2}{\rm TR}[eF\coth(e F s)]\;.\label{xkx}
 \eea
From Eqs.~(\ref{h2}), (\ref{pppp}), and (\ref{xkx}), one finally
obtains
 \beas
 \ch\bbox( x(s), \P(s)\bbox)&=&-x(s)K x(s)+2x(s)K x(0)- x(0)K x(0)\nn\\
 &&-{i\ov 2}{\rm TR}[eF\coth(e F s)]-{e\ov 2}{\rm TR}(\s F)
 +m^2-2z^2b^2\;.
 \eeas
Therefore, Eq.~(\ref{ev}) becomes
 \beas
 i\pl_s\lg x'(s)|x''(0)\rg&=&\biggl[-(x'-x'')K(x'-x'')
 -{i\ov 2}{\rm TR}[eF\coth(e F s)]\nn\\
 &&-{e\ov 2}{\rm TR}(\s F)+m^2-2z^2b^2\biggr]\lg x'(s)|x''(0)\rg\;,
 \eeas 
whose solution is 
 \bea
 \lg x'(s)|x''(0)\rg&=&C(x',x'')s^{-2}{\rm e}^{-L(s)}
 \exp\biggl[-{i\ov 4}(x'-x'')eF\coth(e F s)(x'-x'')\nn\\
 &&+{ie\ov 2}{\rm TR}(\s F)s-i(m^2-2z^2b^2)s\biggr]\;, \label{sol}
 \eea
where
 \beas
 L(s)={1\ov 2}{\rm TR}\ln[(e F s)^{-1}\sinh(e F s)]\;.
 \eeas

From the definition of the operator $\P$ and the first two relations 
in Eq.~(\ref{pppp}), one can find
 \bea
 &&[i\pl' -e A(x')]\lg x'(s)|x''(0)\rg
 =\biggl({1\ov 2}e F[\coth(e F s)+1]
 (x'-x'')+iz\s b \g_5\biggr)\lg x'(s)|x''(0)\rg\;,\nn\\
 &&[-i\pl'' -e A(x'')]\lg x'(s)|x''(0)\rg
 =\biggl({1\ov 2}e F[\coth(e F s)-1]
 (x'-x'')+iz\s b \g_5\biggr)\lg x'(s)|x''(0)\rg\;, \label{2e}
 \eea
from which, in conjunction with Eq.~(\ref{sol}),
the differential equations for $C(x',x'')$
are obtained:
 \beas
 \biggl[i\pl'-e A(x')-{e\ov 2}F(x'-x'')-iz\s b\g_5
 \biggr]C(x',x'')&=&0\;, \nn\\
 \biggl[-i\pl''-e A(x'')+{e\ov 2}F(x'-x'')-iz\s b\g_5
 \biggr]C(x',x'')&=&0\;. 
 \eeas
Two forms of the  solution for $C(x',x'')$ are obtained:
 \bea
 C(x',x'')&=&C_1(x'')\exp\biggl[z(x'-x'')\s b \g_5-ie\int_{x''}^{x'} dx^\m
 \Bigl( A(x)+{1\ov 2}F (x-x'')\Bigr)_\m\biggr]\;, \label{cc1}\\
 C(x',x'')&=&C_2(x')\exp\biggl[z(x'-x'')\s b \g_5+ie\int_{x'}^{x''} dx^\m
 \Bigl( A(x)+{1\ov 2}F (x-x') \Bigr)_\m\biggr]\;,\label{cc2}
 \eea
Since both integrals in Eqs.~(\ref{cc1}) and (\ref{cc2})
are independent of the integration path due to the Stokes' theorem, 
one may choose the integration 
path to be a straight line connecting $x'$ and $x''$,
i.e., $x=x''+t(x'-x'')$. Then, it is readily shown that
 \beas
 &&\int_{x''}^{x'} dx^\m \Bigl(F (x-x'')\Bigr)_\m
 =\int_0^1 dt(x'-x'')F(x'-x'')=0\;,\nn\\
 &&\int_{x'}^{x''} dx^\m \Bigl(F (x-x')\Bigr)_\m
 =\int_1^0 dt(x'-x'')F(x'-x'')(1-t)=0\;.
 \eeas
With these vanishing integrals in mind, the comparison of two
expressions for $C(x',x'')$ in Eqs.~(\ref{cc1}) and (\ref{cc2}) 
leads us to the following conclusion: $C_1(x'')= C_2(x')=\cc$ (constant).
The constant $\cc$ is finally determined by Eq.~(\ref{bc}) as  
 \beas
 \cc=-{i\ov (4\p)^2}\;.
 \eeas
Therefore, the complete form of the transition amplitude is given as
\bea
 \lg x'(s)|x''(0)\rg&=&-{i\ov (4\p)^2}\F(x',x'')s^{-2}{\rm e}^{-L(s)}
 \exp\biggl[-{i\ov 4}(x'-x'')eF\coth(e F s)(x'-x'')\nn\\  
 &&+{ie\ov 2}{\rm TR}(\s F)s-i(m^2-2z^2b^2)s\biggr]\;, \label{ta}
 \eea
where 
 \bea
 \F(x',x'')=\exp\biggl[z(x'-x'')\s b \g_5
 -ie \int_{x''}^{x'} dx^\m  A_\m(x)\biggr]\;. \label{fxx}
 \eea

Now, let us compute ${\rm tr}[ G(x,x')\g_5 \bs]$ in Eq.~(\ref{g1}). 
From Eqs.~(\ref{gd1}), (\ref{gd2}), and (\ref{d2}), we obtain
  \beas 
 {\rm tr}[ G(x,x')\g_5 \bs]&=&{\rm tr}\biggl[(i \g_5 b^\m-\g_5\s^{\m\n}b_\n)
 \int_0^\8 ds \lg x(s)|\P_\m(s)|x'(0)\rg\biggr]\nn\\
 &&+izb^2{\rm tr}\biggl[\int_0^\8 ds \lg x(s)|x'(0)\rg\biggr]\nn\\
 &=&{\rm tr}\biggl[(-i\g_5 b^\m-\g_5\s^{\m\n}b_\n)
 \int_0^\8 ds  \lg x(s)|\P_\m(0)|x'(0)\rg\biggr]\nn\\
 &&+izb^2{\rm tr}\biggl[\int_0^\8 ds \lg x(s)|x'(0)\rg\biggr]\;.
 \eeas
Averaging these two equivalent expressions gives
 \bea 
 {\rm tr}[ G(x,x')\g_5 \bs]&=&{ib^\m\ov 2}{\rm tr}\biggl[
 \g_5\int_0^\8 ds \lg x(s)|[\P_\m(s)-\P_\m(0)]|x'(0)\rg\biggr]\nn\\
 &&-{b_\n\ov 2}{\rm tr}\biggl[\g_5\s^{\m\n}
 \int_0^\8 ds  \lg x(s)|[\P_\m(s)+\P_\m(0)]|x'(0)\rg\biggr]\nn\\
 &&+izb^2{\rm tr}\biggl[\int_0^\8 ds \lg x(s)|x'(0)\rg\biggr]\;. \label{ggb}
 \eea
From the first two relations in Eq.~(\ref{pppp}), we have 
 \bea
 \lg x(s)|\P_\m(s)-\P_\m(0)|x'(0)\rg&=&
 [e F (x-x')]_\m\lg x(s)|x'(0)\rg\;,\nn\\
 \lg x(s)|\P_\m(s)+\P_\m(0)|x'(0)\rg&=&
 [e F\coth(e F s) (x-x')+iz\s b \g_5]_\m\lg x(s)|x'(0)\rg\;.\label{ppm}
 \eea
Using Eqs.~(\ref{ta}) and (\ref{ppm}), we finally rewrite Eq.~(\ref{ggb})
as follows:
\bea 
 {\rm tr}[ G(x,x')\g_5 \bs]
 &=&{1\ov (4\p)^2}\F_0(x,x')\int_0^\8 
 {ds\ov s^2}\,{\rm e}^{-i(m^2-2z^2b^2)s}\,{\rm e}^{-L(s)}\,
 {\rm e}^{-{i\ov 4}(x-x')eF\coth(e F s)(x-x')}\nn\\
 &&\times\biggl\{{1\ov 2}eF_{\m\n}(x-x')^\n b^\m
 {\rm tr}\biggl[\g_5\, {\rm e}^{z(x-x')\s b}\,
 {\rm e}^{{ie\ov 2}{\rm TR}(\s F)s}\biggr]\nn\\ 
 &&+{ie\ov 2}F_{\m\a}[\coth(e F s)]^{\a\b}(x-x')_\b b_\n
 {\rm tr}\biggl[\g_5\s^{\m\n}\,{\rm e}^{z(x-x')\s b}\,
 {\rm e}^{{ie\ov 2}{\rm TR}(\s F)s} \biggr]\nn\\
 &&-{z\ov 2}b^2{\rm tr}\biggl[{\rm e}^{z(x-x')\s b}\,
 {\rm e}^{{ie\ov 2}{\rm TR}(\s F)s} \biggr] \biggr\}\;, \label{ggb2}
 \eea
where $\F_0(x,x')$ is the function $\F(x,x')$ in Eq.~(\ref{fxx}) 
at $z=0$.

The vacuum current vector $\lg J_\m(x)\rg$ associated with 
$b$ is obtained from $\G_{\rm eff}^{(1)}$ by variation of $A^\m (x)$:
\bea
\lg J_\m(x)\rg&=&{\d \G_{\rm eff}^{(1)}\ov \d A^\m (x)}\;.
\eea
Thus, from Eqs.~(\ref{g1}) and (\ref{ggb2}), we have, after deforming 
the integration path of $is$ to the positive real axis,
 \bea
 \lg J_\m(x)\rg
 &=&{ie\ov (4\p)^2}(x-x')_\m\F_0(x,x')\int_0^1 dz\,\int_0^\8 
 {ds\ov s^2}\,\nn\\
 &&\times{\rm e}^{-(m^2-2z^2b^2)s}\,{\rm e}^{-\ell(s)}\,
 {\rm e}^{{1\ov 4}(x-x')eF\cot(e F s)(x-x')}\nn\\
 &&\times\biggl\{-{1\ov 2}eF_{\r\n}(x-x')^\n b^\r
 {\rm tr}\biggl[\g_5\, {\rm e}^{z(x-x')\s b\g_5}\,
 {\rm e}^{{e\ov 2}{\rm TR}(\s F)s}\biggr]\nn\\
 &&+{e\ov 2}F_{\r\a}[\cot(e F s)]^{\a\b}(x-x')_\b b_\n
 {\rm tr}\biggl[\g_5\s^{\r\n}\,{\rm e}^{z(x-x')\s b\g_5}\,
 {\rm e}^{{e\ov 2}{\rm TR}(\s F)s} \biggr]\nn\\ 
 &&+{z\ov 2}b^2{\rm tr}\biggl[{\rm e}^{z(x-x')\s b\g_5}\,
 {\rm e}^{{e\ov 2}{\rm TR}(\s F)s} \biggr]
 \biggr\} _{x'\rar x}\;, \label{j}
\eea
where
 \beas
 \ell(s)&=&{1\ov 2}{\rm TR}\ln[(e F s)^{-1}\sin(e F s)]\;.
 \eeas
Using the eigenvalue technique \cite{sc}, 
${\rm e}^{-\ell(s)}$ is determined as 
 \bea
 {\rm e}^{-\ell(s)}=
 {(es)^2\cg\ov {\rm Im\,}\cosh(esX)}\;, \label{ep}
 \eea
where
 \beas
 X=\sqrt{2 (\cf+i\cg)}\;,~~~
 \cf={1\ov 4}F_{\m\n}F^{\m\n}=-{1\ov 2}
 (\mbox{\bf E}^2-\mbox{\bf B}^2)\;,~~~
 \cg={1\ov 4}F_{\m\n}\~{F}^{\m\n}=-\mbox{\bf E}\cdot \mbox{\bf B}
 \eeas 
with $\~{F}^{\m\n}={1\ov 2}\e^{\m\n\l\r}F_{\l\r}$.
From the matrix decomposition formulas
 \beas
 &&{\rm e}^{z(x-x')\s b\g_5}=a_1+a_2 (x-x')\s b\g_5\;,\nn\\
 &&{\rm e}^{{e\ov 2}{\rm TR}(\s F)s}=c_1+c_2{\rm TR}(\s F)+
 ic_3\g_5+ic_4{\rm TR}(\s F)\g_5\;,
 \eeas 
where
 \beas
 &&a_1=\sum_{N=0}^\8 {z^{2N}\ov (2N)!}[(x-x')^2b^2-
 \bbox((x-x')\cdot b\bbox)^2]^N\;,\nn\\
 &&a_2=\sum_{N=0}^\8 {z^{2N+1}\ov (2N+1)!}[(x-x')^2b^2-
 \bbox((x-x')\cdot b\bbox)^2]^N\;,\nn\\
 &&c_1={\rm Re}\,\cosh(esX)\;,~~~~
 c_2={\rm Re}\,[\sinh(esX)/(2X)]\;,\nn\\~~~~
 &&c_3={\rm Im}\,\cosh(esX)\;,~~~~
 c_4={\rm Im}\,[\sinh(esX)/(2X)]\;,
 \eeas
the trace quantities in the integrand of Eq.~(\ref{j}) are 
evaluated as follows:
 \bea
 &&{\rm tr}\biggl[{\rm e}^{z(x-x')\s b\g_5}\,
 {\rm e}^{{e\ov 2}{\rm TR}(\s F)s} \biggr]
  =4a_1c_1-8ia_2c_2(x-x')\~{F} b-8ia_2c_4(x-x')Fb\;,\nn\\
 &&{\rm tr}\biggl[\g_5\, {\rm e}^{z(x-x')\s b\g_5}\,
 {\rm e}^{{e\ov 2}{\rm TR}(\s F)s}\biggr]
  =4ia_1c_3-8a_2c_2(x-x')Fb+8a_2c_4(x-x')\~{F}b\;,\nn\\
 &&{\rm tr}\biggl[\g_5\s^{\r\n}\,{\rm e}^{z(x-x')\s b\g_5}\,
 {\rm e}^{{e\ov 2}{\rm TR}(\s F)s} \biggr]
 =-8ia_1 \Bigl(c_2 \~{F}^{\r\n}+c_4 F^{\r\n}\Bigr)\nn\\
 &&~~~~~~~~~~~~~~+4a_2 \Bigl(c_1[(x-x')^\r b^\n
 -(x-x')^\n b^\r]+ic_3(x-x')_\a b_\b \e^{\a\b\r\n}\Bigr)\nn\\
 &&~~~~~~~~~~~~~~+8ia_2\Bigl(c_2[b^\r\bbox((x-x')F\bbox)^\n-b^\n
            \bbox((x-x')F\bbox)^\r\nn\\
 &&~~~~~~~~~~~~~~+(x-x')^\n(bF)^\r-(x-x')^\r(bF)^\n]\nn\\
 &&~~~~~~~~~~~~~~+4ic_4[\e^{\r\n\a\b}(x-x')_\a(bF)_\b
 -\e^{\r\n\a\b}b_\a\bbox((x-x')F\bbox)_\b\nn\\
 &&~~~~~~~~~~~~~~-b^\r\bbox((x-x')\~{F}\bbox)^\n
 +b^\n\bbox((x-x')\~{F}\bbox)^\r\nn\\
 &&~~~~~~~~~~~~~~+(x-x')^\r(b\~{F})^\n
 -(x-x')^\n(b\~{F})^\r]\Bigr)\;. \label{tr}
 \eea

Now substituting Eqs.~(\ref{ep}) and (\ref{tr}) into Eq.~(\ref{j}) and 
discarding terms that do not contribute to the Chern-simons current,
we obtain 
 \bea
 \lg J_\m(x)\rg
 &=&{e^3\cg\ov 4\p^2}(x-x')_\m\F_0(x,x')\int_0^1 dz\,\int_0^\8 
 ds\,{1\ov {\rm Im}\,\cosh(esX)}\nn\\
 &&\times{\rm e}^{-(m^2-2z^2b^2)s}\,
 {\rm e}^{{1\ov 4}(x-x')eF\coth(e F s)(x-x')}\nn\\
 &&\times\biggl\{a_1 c_2eF_{\r\a}[\cot(e F s)]^{\a\b}(x-x')_\b b_\n
 \~{F}^{\r\n}+a_2c_2zb^2 (x-x')_\r\~{F}^{\r\n}b_\n 
 \biggr\} _{x'\rar x}\;. \label{j1}
 \eea
Thus, the current $\lg J_\m(x)\rg$ of Eq.~(\ref{j1})
is cast, in the leading order approximation in the coupling constant 
$e$, to the following form:\footnote{This approximation matches with the 
conventional Feynman diagram calculation of the {\em one-loop} 
vacuum polarization tensor.}
 \bea
 \lg J_\m(x)\rg
 &=&{e^2\ov 8\p^2}\int_0^1 dz
 \biggl[(x-x')_\m(x-x')_\r\F_0(x,x')
 \~{F}^{\r\n} b_\n\nn\\
 &&\times\int_0^\8 ds\biggl({a_1\ov s^2}
 +{a_2 z b^2\ov s}\biggr){\rm e}^{-(m^2-2z^2b^2)s+(x-x')^2/(4s)}
 \biggr]_{x'\rar x}\;.  \label{j2}
\eea

In the light of $b$-perturbation theory, the insertions of $\bs \g_5$
are revealed best via the expansion of the factor
${\rm e}^{2z^2 b^2 s}$ [in Eq.~(\ref{j2})] in powers of its exponent. 
Then using an integration formula 
 \beas
 \int_0^\8 ds s^{n-2}\,{\rm e}^{-sy-1/s}=2 y^{(1-n)/2}
 K_{n-1}\Bigl(2 y^{1/2}\Bigr)\;,
 \eeas
with $K_n(x)$ being the modified Bessel function of order $n$,
the Chern-Simons current $\lg J_\m(x)\rg$ in Eq.~(\ref{j2})
is evaluated as follows:
 \bea
 \lg J_\m(x)\rg&=&{e^2\ov 2\p^2}\int_0^1 dz\biggl[-\F_0(x,x')
 {(x-x')_\m(x-x')_\r\ov (x-x')^2}\,b_\n\~{F}^{\r\n}\nn\\
 &&\times \sum_{N=0}^\8 {z^{2N}\ov (2N)!}\biggl(1
 +{z^2\ov 2N+1}{b^2\ov m^2}[-m^2 (x-x')^2]^{1/2}\biggr)\nn\\
 &&\times \biggl\{(x-x')^2b^2
 -{(x-x')_\a(x-x')_\b\ov (x-x')^2}(x-x')^2 b^\a b^\b\biggr\}^N\nn\\
 &&\times \sum_{n=0}^\8 {z^{2n}\ov n!}\biggl({b^2\ov m^2}\biggr)^n\,
 [-m^2(x-x')^2]^{(n+1)/2}\,
 K_{n-1}\Bigl([-m^2(x-x')^2]^{1/2}\Bigr)
 \biggr]_{x'\rar x}\;.  \label{j4}
 \eea
Using the following formulas 
 \bea
 &&\lim_{x'\rar x}\F_0(x,x')=1\;, \nn\\
 &&\lim_{x'\rar x}[-m^2(x-x')^2]^{1/2}\,
 K_{-1}\Bigl([-m^2(x-x')^2]^{1/2}\Bigr)=1\;, \nn\\ 
 &&\lim_{x'\rar x}[-m^2(x-x')^2]^{(n+1)/2}\,
 K_{n-1}\Bigl([-m^2(x-x')^2]^{1/2}\Bigr)=0\;,~~~~n=1,2,3,\cdots \nn\\
 &&\lim_{x'\rar x}{(x-x')_\m(x-x')_\n\ov (x-x')^2}=c g_{\m\n}\;,
 ~~~(\mbox{$c$ being an arbitrary contant}) \label{cgmn}
 \eea
it is not difficult to see that  
each higher order term separately vanishes.
Thus, we finally arrive at
 \beas
 \lg J_\m(x)\rg&=&-{ce^2\ov 2\p^2}\~{F}_{\m\n}b^\n
 \int_0^1 dz\;,  
 \eeas
and equivalently
 \beas
 \G_{\rm CS}={ce^2\ov 4\p^2}
 \int d^4 x \,\e^{\m\n\l\r}b_\m A_\n F_{\l\r}
 \int_0^1 dz\;, 
 \eeas
which means that the all-order-in-$b$ calculation does not alter
the linear-in-$b$ result and thus the functiom $f(b^2/m^2)$
in Eq.~(\ref{f}) vanishes. This complete our calculation.

We understand \cite{jm3,pv} the identity of the lowest-order calculation
with the all-order calculation by following an argument of Coleman
and Glashow \cite{cg}. Since in the expansion of 
the $b$-dependent vacuum polarization amplitude in powers of $b$ 
($b$-perturbation theory), all
except the first order are free of linear divergences. Hence there is
no ambiguity in evaluating higher order graphs. Momentarily let each of 
the two photons carry different momenta, say $p_1$ and $p_2$ (this means
that the chiral inserions of $\bs \g_5$ carry non-zero momentum); 
Applying the Ward identity to this vacuum polarization amplitude, 
one finds that the amplitude is $O(p_1)$ and $O(p_2)$ at the same time, 
i.e., it is $O(p_1 p_2)$; now go to equal momenta, $p_1=p_2=p$, 
and observe that the amplitude 
must be $O(p^2)$. The Chern-Simons term one is seeking is $O(p)$; hence
all these higher-order graphs do not contribute. 

In summary, we have used the the Fock-Schwinger proper time method
to calculate the induced Chern-Simons term arising from 
the Lorentz- and CPT-violating sector of quantum electrodynamics
with a $b_\m \yb \g^\m \g_5 \y$ term. Although there is an ambiguity
in the calculation as mentioned in footnote 2, 
our all-order-in-$b$ result coincides, 
{\em independently of this ambiguity}, with the linear-in-$b$ result.

\acknowledgments
This work was supported by a Korea Research Foundation grant 
(KRF-2000-015-DP0066).


\end{document}